\documentclass[conference]{IEEEtran}
\IEEEoverridecommandlockouts
\usepackage{cite}
\usepackage{amsmath,amssymb,amsfonts}
\usepackage{algorithmic}
\usepackage{graphicx}
\usepackage{textcomp}
\usepackage{xcolor}
\def\BibTeX{{\rm B\kern-.05em{\sc i\kern-.025em b}\kern-.08em
    T\kern-.1667em\lower.7ex\hbox{E}\kern-.125emX}}
\begin{document}

\title{Blockprint Accuracy Study\\
\thanks{Grant FY23-1321 from the Ethereum Foundation.}
}

\author{\IEEEauthorblockN{Santiago Somoza}
\IEEEauthorblockA{\textit{MigaLabs} \\
Buenos Aires, Argentina \\
santiagosomoza1@gmail.com}
\and
\IEEEauthorblockN{Tarun Mohandas-Daryanani}
\IEEEauthorblockA{\textit{MigaLabs} \\
Barcelona, Spain \\
tarun.mdaryanani@gmail.com}
\and
\IEEEauthorblockN{Leonardo Bautista-Gomez}
\IEEEauthorblockA{\textit{MigaLabs} \\
Barcelona, Spain \\
leobago@protonmail.com}
}

\maketitle

\begin{abstract}
Blockprint, a tool for assessing client diversity on the Ethereum beacon chain, is essential for analyzing decentralization. This paper details experiments conducted at MigaLabs to enhance Blockprint's accuracy, evaluating various configurations for the K-Nearest Neighbors (KNN) classifier and exploring the Multi-Layer Perceptron (MLP) classifier as a proposed alternative. Findings suggest that the MLP classifier generally achieves superior accuracy with a smaller training dataset. The study revealed that clients running in different modes, especially those subscribed to all subnets, impact attestation inclusion differently, leading to proposed methods for mitigating the decline in model accuracy. Consequently, the recommendation is to employ an MLP model trained with a combined dataset of slots from both default and “subscribed to all subnets" client configurations.
\end{abstract}


\section{Introduction}
Blockchains, have revolutionized various industries by enabling the development of decentralized applications. These operate without a central authority, relying on a consensus mechanism to guarantee the validity of transactions. Ethereum \cite{wood2014ethereum}, a prominent blockchain platform, utilizes a Proof-of-Stake \cite{buterin2022proof} based system where validators secure the network by staking their own cryptocurrency, Ether.

Within this context, achieving a high degree of decentralization is paramount \cite{brown2024exploring}. It ensures that no single entity wields excessive control over the validation process. This is where consensus clients and client diversity \cite{grandjean2023ethereum} come into play. Clients are software programs used by validators to participate in Ethereum's consensus mechanism. Having a diverse client set mitigates potential vulnerabilities and bolsters the network's resistance to manipulation.

Blockprint \cite{blockprint} is a tool for measuring client diversity. It relies on a machine learning algorithm trained to classify the client that produced a block based on a list of features extracted from the slot rewards data. 
By analyzing client distribution, Blockprint provides invaluable insights into the decentralization of the Ethereum network. 




This study aimed to enhance Blockprint's accuracy by conducting a comprehensive analysis of its underlying machine learning model. To achieve this, we collected a diverse dataset comprising reward data from six leading consensus clients \cite{clientdiversityorg}, each operating under three different configurations. This dataset simulated real-world conditions, providing a robust foundation for model evaluation.

We began by scrutinizing Blockprint's existing KNN classifier, experimenting with various K values and training set sizes. Subsequently, we introduced the MLP classifier as an alternative, exploring different architectures and training parameters. All models were evaluated across the three client configurations to assess their performance under varying conditions and potential feature variations.

\section{Methodology} \label{methodology}
We ran the six most used consensus clients (Lighthouse, Prysm, Teku, Nimbus, Lodestar and Grandine) connected to a Nethermind execution client and triggered simulated block production for 25,000 slots on each of the clients with three different configurations (default, subscribed to all subnets and with proposer flags activated or utilizing block proposer API calls depending on the client). This resulted in 450,000 total blocks stored. 

Following the acquisition of rewards data, we employed Blockprint models to assess the classification accuracy achieved under various parameter combinations and classifier configurations. To evaluate the accuracy of the constructed models, a standard data splitting technique was implemented. This process divides the dataset into training and testing sets. The training set is used to train the model, while the testing set, comprised of unseen data, is employed for performance evaluation. To mitigate the influence of data point allocation on the evaluation process, cross-validation \cite{peterson2009k} was utilized.

\section{Experiments Conducted} \label{experiments}


\subsection{Classifier Parameter optimization}
To assess the impact of individual parameters on model accuracy, we experimented with different configurations of a KNN model using a fixed dataset of 18,000 reward data points (3,000 per client). Our findings indicate that the optimal value for the K parameter lies within the range of 9 to 14, all achieving a peak accuracy of around 91.5\%. To determine the optimal training data size, we fixed K at 9 and gradually increased the dataset size from 6,000 to 120,000 while maintaining equal reward data distribution across clients. Results show that model accuracy improved with increasing dataset size, eventually saturating beyond 60,000 reward data points.

We introduced the MLP classifier as an alternative to the KNN. Utilizing two hidden layers of sizes $(391, 870)$ chosen arbitrarily, it outperformed KNN significantly using the same dataset sizes as in the KNN analysis. The MLP achieved a 3.5\% accuracy advantage over KNN with the smallest dataset. While MLP accuracy improved with increasing dataset size, the gains diminished, plateauing around 95.5\% with 102,000 data points. This suggests that expanding the dataset beyond 66,000 points might not yield substantial benefits, similar to the KNN findings.


The number of possible configurations for hidden layers in a neural network is unbounded by any practical limit. It’s best to experiment with the dataset to be classified using a fine-tuning algorithm like GridSearch \cite{bergstra2011algorithms} or RandomSearch \cite{bergstra2012random}. We ran RandomSearch with the number of layers ranging from 1 to 10 and the individual hidden layer sizes ranging from 100 to 2000, trained with 18,000 slot rewards data. We observed that utilizing more than 2 layers did not improve the accuracy, having models with 1-2 layers of sizes 800-1200 among the best performing. 

\subsection{Different Client modes}
In the consensus protocol of Ethereum, a node is required to be subscribed to the number of attestation subnets (attnets) equivalent to the number of validators that it will be hosting (with a maximum of 64). Since PR 3312 \cite{PR3312} all nodes are required to subscribe to two attnets. The more attnets the node is connected to, the more processing and bandwidth requirements are for running the node. We observed that subscribing to all subnets led some clients to include more redundant attestations, modifying the features related to those. It is worth noting that future updates to the protocol will require nodes to subscribe to more attnets. This will impact the accuracy of Blockprint requiring retraining the model.

At MigaLabs, we track the amount of attestation subnets that nodes are subscribed to in the network. Only 5.5\% of the nodes are subscribed to all subnets. However, we can’t get any information on which of these are validators and which are non-validators. We are unaware of any report that guesses the percentage of validator nodes that are subscribed to all attnets, but after having experimented, we believe it might be possible to estimate this with decent accuracy using a modified version of blockprint. 

We gathered blocks produced by nodes running “subscribed to all subnets” and tested our models with these. In the case of an MLP model, the accuracy with the default dataset was 94.6\%, dropping to 87.6\% when tested with the all subnets dataset. For another KNN model, similar to the MLP, the accuracy dropped from 93\% to 89\%. In both cases, the models fail to classify correctly nimbus rewards data. This is because being subscribed to all subnets changes the attestation inclusion for this client, causing the classifier to confuse it with others like Grandine. The accuracy also drops in at least 2\% for Grandine and Prysm. Having observed that some clients fail to be classified with the model trained on the default dataset, we considered either merging both datasets or treating the different modes as different classes (e.g. \textit{Lighthouse} and \textit{Lighthouse\_all\_subnets}), using 12 classes instead of 6.

Merging both datasets with equal parts of each achieved very good results (over 95\% accuracy, with an MLP classifier even reaching 96.1\%). The increase in accuracy compared to the model that only uses the default dataset is due to the clients being more easily classifiable when running with the flag enabled. The second alternative shows evidence that some clients have their features modified significantly by running with the flag on. Training an MLP classifier with this configuration using 36,000 rewards data objects (3,000 per client per mode) resulted in 75\% accuracy. We observed that some classes are confused with others depending on the mode the client was run.

Another mode of running the clients is utilizing proposer flags, which are supported on some of the clients and calling the \textit{/validator/prepare\_beacon\_proposer} API endpoint on the rest in the middle of the previous epoch to signal the execution client to start preparing from beforehand the block proposal. We found no noticeable difference in the way that blocks are produced, achieving the same accuracy on models trained with the default dataset. 
\section{Conclusions} \label{conclusions}
Our analysis of Blockprint's KNN classifier explored various configurations to optimize performance. We proposed the use of MLP, which emerged as a superior classifier, demonstrating improved accuracy compared to KNN while requiring a much smaller training dataset. Additionally, we observed that clients subscribed to all subnets have their attestation inclusion methods modified which impacts the accuracy of the model if not trained accordingly. To address this, we propose utilizing a mixed dataset and continuous model monitoring to maintain optimal performance. Based on our findings, we recommend employing an MLP model trained on a combined dataset of default and “subscribed to all subnets" client configurations.
\section*{Acknowledgments} \label{acknowledgments}
The authors thank Michael Sproul for his insightful feedback. This research was supported by the Ethereum Foundation (Grant FY23-1321).%
\bibliographystyle{IEEEtran}
\bibliography{IEEEabrv,mybibfile}

\end{document}